\documentclass[unnumsec,webpdf,contemporary,large]{oup-authoring-template}%
\usepackage{balance}

\graphicspath{{Fig/}}

\theoremstyle{thmstyleone}%
\theoremstyle{thmstyletwo}%
\theoremstyle{thmstylethree}%

\usepackage{hyperref}
\hypersetup{
    colorlinks = true,
    linkcolor = blue,
    urlcolor = blue
}

\begin{document}

\journaltitle{Journal Title Here}
\DOI{DOI HERE}
\copyrightyear{2022}
\pubyear{2019}
\access{Advance Access Publication Date: Day Month Year}
\appnotes{Paper}

\firstpage{1}

%\subtitle{Subject Section}

\title[FastDup]{FastDup: a scalable duplicate marking tool using speculation-and-test mechanism}

\author[1,2]{Zhonghai Zhang}
\author[3]{Yewen Li}
\author[1]{Ke Meng}
\author[1]{Chunming Zhang}
\author[1,2,$\ast$]{Guangming Tan}

\authormark{Author Name et al.}

\address[1]{
    % \orgdiv{Department of Computer Science},
    \orgname{Institute of Computing Technology},
    \orgaddress{
        \street{No. 6 Kexueyuan South Road},
        \postcode{100190},
        \state{Haidian District},
        \country{Beijing, China}
    }
}

\address[2]{
    \orgdiv{School of Computer Science},
    \orgname{University of Chinese Academy of Sciences},
    \orgaddress{
        \street{No. 19A Yuquan Road},
        \postcode{100049},
        \state{Haidian District},
        \country{Beijing, China}
    }
}

\address[3]{
    \orgname{The Hong Kong University of Science and Technology},
    \orgaddress{
        \street{Sai Kung District},
        \postcode{999077},
        \state{New Territories},
        \country{Hong Kong, China}
    }
}

\corresp[$\ast$]{Corresponding author. \href{email:tgm@ict.ac.cn}{tgm@ict.ac.cn}}

\received{Date}{0}{Year}
\revised{Date}{0}{Year}
\accepted{Date}{0}{Year}

%\editor{Associate Editor: Name}

%\abstract{
%\textbf{Motivation:} .\\
%\textbf{Results:} .\\
%\textbf{Availability:} .\\
%\textbf{Contact:} \href{name@email.com}{name@email.com}\\
%\textbf{Supplementary information:} Supplementary data are available at \textit{Journal Name}
%online.}

\abstract{
\textbf{Summary:}
Duplicate marking is a critical preprocessing step in gene sequence analysis to flag redundant reads arising from polymerase chain reaction(PCR) amplification and sequencing artifacts. Although Picard MarkDuplicates is widely recognized as the gold-standard tool, its single-threaded implementation and reliance on global sorting result in significant computational and resource overhead, limiting its efficiency on large-scale datasets. Here, we introduce FastDup: a high-performance, scalable solution that follows the speculation-and-test mechanism. FastDup achieves up to 20x throughput speedup and guarantees 100\% identical output compared to Picard MarkDuplicates.\\
\textbf{Availability and implementation:}
FastDup is a C++ program available from GitHub \url{https://github.com/zzhofict/FastDup.git} under the MIT license.}


\maketitle

\section{Introduction}
% 介绍冗余的来源, 以及标记冗余的重要性
Duplicate reads are a pervasive issue in next-generation sequencing (NGS) data and arise primarily due to two sources: PCR amplification during library preparation and sequencing artifacts such as optical duplicates. These duplicate reads can skew downstream analyses by inflating read counts, distorting allele frequency estimates, and potentially leading to false positive variant calls. Consequently, identifying the duplicate reads is a critical preprocessing step in gene sequence analysis.

% 介绍当前的标记冗余的工具
% 分开介绍，先介绍传统的picard markduplicates，再介绍流式处理的streammd等，介绍他们的原理，优势和缺点

%GATK MarkDuplicatesSpark和sambamba用并行的方式实现了相同的冗余检测算法，提升了一定的性能，但是他们都依赖与全局排序，这导致他们占用大量的内存资源来保存中间结果，或者将中间结果放入硬盘，这样产生大量的中间文件，增加了IO，这都影响了性能的提升
Several tools have been developed to address this problem. Among these, Picard MarkDuplicates~\cite{Picard_Toolkit} has long been recognized as the gold standard due to its robust duplicate identification algorithm. However, its reliance on position-sorted data and single-threaded implementation limits its performance, especially for large-scale datasets. GATK MarkDuplicatesSpark~\cite{GATK_2020} and Sambamba~\cite{sambamba_2015} use the same duplicate identification algorithm as Picard MarkDuplicates but achieve better performance by leveraging parallel processing. However, they still depend on the global sorting results of genomic coordinates of all reads, which requires either a significant memory footprint or writing intermediate results to disk, thereby constraining their overall efficiency.

Streaming-based approaches, such as Samblaster~\cite{faust_samblaster_2014} and streammd~\cite{leonard_streammd_2023}, have been proposed to accelerate duplicate marking by eliminating the need for position-sorted reads and enabling direct piping after the mapping process. Samblaster employs a hash-based structure, while streammd utilizes a Bloom filter to identify duplicates in real time. However, both methods simplify duplicate marking by retaining only the first read and treating all subsequent reads with the same genomic coordinate as duplicates, a strategy that can compromise accuracy in scenarios where more nuanced differentiation is required.

% 引出FastDup工作
To overcome these limitations, we present FastDup, a high-performance duplicate marking tool that combines the robustness of Picard’s duplicate identification algorithm with innovative performance optimizations. FastDup employs a novel speculation-and-test mechanism, wherein potential duplicates are speculated in parallel and subsequently verified in a streamlined testing phase. This approach is integrated into a fully parallelized pipeline that avoids global sorting while maintaining identical accuracy compared to Picard MarkDuplicates~\cite{Picard_Toolkit}. Experiment results demonstrate that FastDup achieves over a 20× speedup compared to traditional methods, offering a scalable and efficient solution for high-throughput sequencing workflows.

\section{Description}
\begin{figure*}[t]
	\centering
	\includegraphics[width=0.99\textwidth]{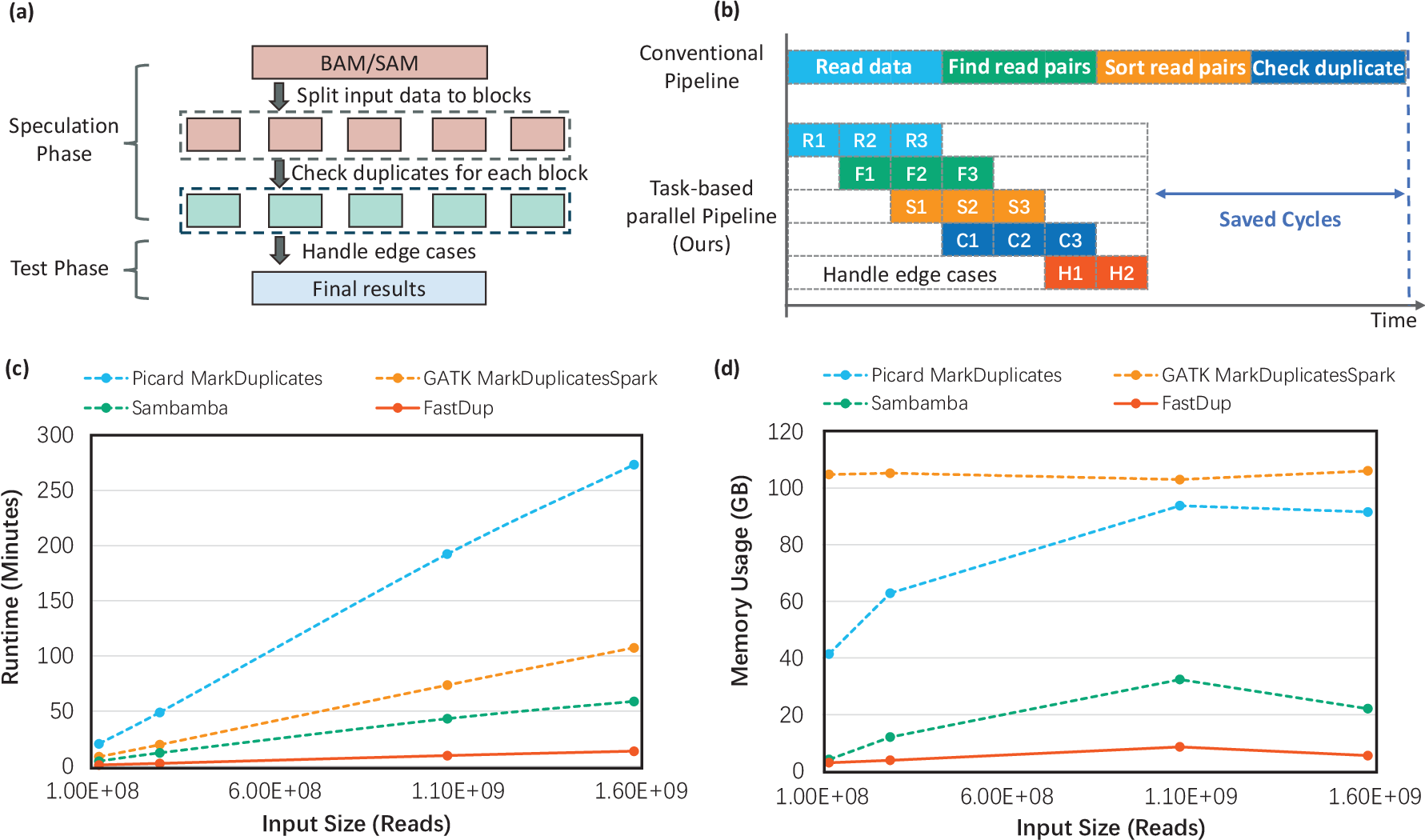}
    \caption{(a) The workflow of speculation-and-test strategy in FastDup. (b) Comparison between conventional pipeline and task-based parallel pipeline in FastDup. (c) Picard MarkDuplicates, GATK MarkDuplicatesSpark, Sambamba and FastDup runtime for four input sizes. (d) Picard MarkDuplicates, GATK MarkDuplicatesSpark, Sambamba and FastDup peak memory usage for four input sizes.}
	\label{fig:optimize-and-result}
\end{figure*}
% 介绍FastDup的实现和特点
FastDup is implemented in C++ and leverages the htslib~\cite{bonfield_htslib_2021} for high-performance parsing of SAM/BAM files. To maximize I/O efficiency, it employs an asynchronous I/O pipeline that decouples file decompression/compression from computational tasks, enabling simultaneous multi-threaded read and write operations through lock-free queues. Similar to Picard MarkDuplicates~\cite{Picard_Toolkit}, FastDup adopts a two-pass duplicate marking strategy to accurately identify all duplicates while retaining the highest-scoring read pairs based on the same scoring system.

%详细介绍FastDup的Speculation-and-test机制
FastDup requires coordinate-sorted SAM/BAM files, where reads are ordered by their mapped genomic positions. However, due to read clipping during alignment, the actual positions used for duplicate detection may differ from the recorded mapping positions. Traditional tools like Picard address this by globally re-sorting reads, creating a computational bottleneck. FastDup circumvents global sorting through a novel speculation-and-test workflow as shown in Fig~\ref{fig:optimize-and-result}(a). This approach is based on the following two observations:
i) Clipping is not common in the mapped reads, which means most reads retain original mapping positions for duplicate detection.
ii) Read pairs typically reside near each other in coordinate-sorted files. The workflow proceeds as follows:
i) Speculation Phase: Split the input into independent data blocks and process each block in parallel, assuming no inter-block dependencies. Identify duplicates within each block using inferred fragment positions.
ii). Test Phase: Resolve edge cases where read pairs span block boundaries by checking neighboring blocks and correct misclassified duplicates using a lightweight dependency graph. This approach eliminates global sorting while maintaining accuracy.

% 为了进一步优化性能，采用了并行流水线策略
To further improve performance, FastDup employs a parallel pipeline architecture that divides duplicate identification into five key tasks, each executed in parallel as shown in Fig~\ref{fig:optimize-and-result}(b):
i) Data Reading and Decompression: Multi-threaded reading and decompression of data blocks.
ii) Signature Generation and Pairing: Creating unique signatures for each read and identifying read pairs within each block.
iii) Sorting: Sorting read signatures for both paired and unpaired reads.
iv) Intra-block Duplicate Identification: Detecting duplicate reads within individual data blocks.
v) Inter-block Dependency Resolution: Resolving duplicate dependencies across neighboring blocks to ensure accuracy.

Additionally, FastDup supports duplicate classification, distinguishing between PCR duplicates and optical duplicates. It also generates a metrics file containing detailed statistics on duplicate occurrences, providing valuable insights for downstream analysis.
\section{Results}

\begin{table}[htbp]
\vspace{1em}
\centering
\small
\caption{Four experimental human datasets~\cite{NCBI2025}.}
\label{tab:datasets}
\begin{tabular}{ccccc}
\toprule
\textbf{Source} & \textbf{Type} & \textbf{Length} & \textbf{\#Reads} & \textbf{\#Duplicates} \\
\midrule
\href{https://trace.ncbi.nlm.nih.gov/Traces/?view=run_browser&acc=SRR25735654&display=metadata}{SRR25735654}     & WES           & 145 bp              & 118,205,082           & 13,019,231      \\
\href{https://trace.ncbi.nlm.nih.gov/Traces/?view=run_browser&acc=SRR8381428&display=metadata}{SRR8381428}      & WES           & 148 bp              & 283,703,444           & 59,461,068      \\
\href{https://trace.ncbi.nlm.nih.gov/Traces/?view=run_browser&acc=SRR25735656&display=metadata}{SRR25735656}     & WGS           & 151 bp              & 1,069,743,600          & 180,186,251    \\
\href{https://trace.ncbi.nlm.nih.gov/Traces/?view=run_browser&acc=ERR194147&display=metadata}{ERR194147}       & WGS           & 101 bp              & 1,582,771,014          & 23,855,846     \\
\hline
\end{tabular}
\end{table}

To evaluate the correctness and performance improvements of FastDup, we compare it against three widely used tools: Picard MarkDuplicates~\cite{Picard_Toolkit}, GATK MarkDuplicatesSpark~\cite{GATK_2020}, and Sambamba~\cite{sambamba_2015}. For each tool, we recorded runtime and peak memory usage across multiple datasets. All input BAM files are publicly available through the NCBI~\cite{NCBI2025} Sequence Read Archive (SRA).

Table 1 summarizes the characteristics of the four datasets used in the evaluation, which include both whole-exome sequencing (WES) and whole-genome sequencing (WGS) data. The number of reads per dataset ranges from 100 million to 1.58 billion. These BAM files were generated using the BWA-MEM aligner. All tools, including FastDup, produced identical duplicate marking results, with the total number of marked duplicates shown in the final column of Table 1.

Fig~\ref{fig:optimize-and-result}(c) presents the wall time required by each tool to complete duplicate marking across the four datasets. The results demonstrate that: i) FastDup achieves an average speedup of 20.13×, 8.03×, and 4.56× over Picard MarkDuplicates, GATK MarkDuplicatesSpark, and Sambamba, respectively. ii) FastDup maintains stable performance across datasets of varying sizes, showing scalability independent of input volume.

Fig~\ref{fig:optimize-and-result}(d) shows the peak memory usage of all tools. The observations are as follows:
i) GATK MarkDuplicatesSpark exhibits the highest memory usage, requiring large memory allocations regardless of dataset size.
ii) Picard MarkDuplicates and Sambamba use less memory, but their memory consumption scales with the input size.
iii) FastDup consistently consumes the least memory among all tools, even when processing the largest datasets.

All experiments were conducted on a 32-core AMD server equipped with an AMD 3970X processor and 128 GB of RAM. Picard MarkDuplicates and GATK MarkDuplicatesSpark were executed using OpenJDK 19.0.2, with a maximum heap size set via -Xmx100G. All input and output files were in BAM format. For tools supporting multithreading, the number of threads was set to 32.

\section{Discussion}
Our evaluation shows that FastDup produces identical results to Picard MarkDuplicates while achieving up to a 20-fold increase in performance. This improvement stems from FastDup’s ability to exploit the structure of coordinate-sorted BAM files—specifically, the observation that most paired reads are located near each other and are approximately sorted by position. As a result, FastDup eliminates the need for expensive global sorting operations. This insight enables finer-grained parallelism: FastDup divides the input into small, independent data blocks, processes each in parallel using a speculation-and-test strategy, and subsequently resolves cross-block dependencies. The use of a parallel pipeline architecture further enhances throughput and overall efficiency.

Picard MarkDuplicates~\cite{Picard_Toolkit}, GATK MarkDuplicatesSpark~\cite{GATK_2020}, and Sambamba~\cite{sambamba_2015} are widely used tools for duplicate marking. While they offer accurate results, all three rely heavily on global coordinate sorting, which leads to significant memory usage. When memory is insufficient, intermediate data must be written to disk, further impacting performance. Additionally, their parallel scalability is limited, making them less suitable for high-throughput or large-scale datasets. These challenges are common limitations across traditional duplicate marking tools.

In contrast to traditional tools, streaming-based approaches such as Samblaster~\cite{faust_samblaster_2014} and streammd~\cite{leonard_streammd_2023} offer real-time duplicate marking by integrating into the mapping pipeline. They utilize lightweight data structures like hash tables or Bloom filters to detect duplicates on the fly. However, these methods do not offer the same level of precision: they may miss certain duplicates or retain suboptimal reads, as they lack access to global sorting information. While faster than traditional tools, their reduced accuracy limits their suitability for high-precision workflows.

FastDup combines the accuracy of traditional tools with the performance benefits of streaming approaches. Its speculation-and-test mechanism allows for fine-grained parallelism while maintaining exact duplicate marking results identical to those of Picard MarkDuplicates~\cite{Picard_Toolkit}. However, it currently only supports coordinate-sorted SAM/BAM input files. FastDup supports seamless integration into the GATK~\cite{GATK_2020} genomic analysis pipeline.

\section*{Acknowledgments}

We acknowledge ZheYuan Technology and Westen Institute of Computing Technology for generously providing computational resources to support this work.

%\section*{Supplementary data}
%Supplementary data are available at Bioinformatics online.

\section*{Conflict of interest}

None declared.

\section*{Funding}

This work was supported in part by National Natural Science Foundation of China(62032023) and National Science Fund for Distinguished Young Scholars(T2125013).

\section*{Data availability}

WES and WGS data used in the performance evaluation is publicly available through the NCBI Sequence Read Archive (SRA). The SRA accession IDs are listed in Table 1.

\bibliographystyle{ACM-Reference-Format}
\balance
\bibliography{oup-authoring-template}

\end{document}